\documentclass{article}

    \PassOptionsToPackage{numbers, compress}{natbib}

\usepackage[final]{neurips_2021}

\usepackage[utf8]{inputenc} 
\usepackage[T1]{fontenc}    
\usepackage{hyperref}       
\usepackage{url}            
\usepackage{booktabs}       
\usepackage{amsfonts}       
\usepackage{nicefrac}       
\usepackage{microtype}      
\usepackage{xcolor}         
\usepackage[linesnumbered,ruled,vlined]{algorithm2e}

\usepackage{amssymb}
\usepackage{latexsym}
\usepackage{amsmath,amssymb}
\usepackage{graphicx}
\usepackage{multirow}
\usepackage{adjustbox}
\usepackage{makecell}
\usepackage[nottoc]{tocbibind}

\usepackage{mathrsfs}

\newcommand{\Bc}{{\mathcal{B}}}
\newcommand{\Cc}{{\mathcal{C}}}
\newcommand{\Hc}{{\mathcal{H}}}
\newcommand{\Tc}{{\mathcal{T}}}
\newcommand{\add}[1] {\textcolor{black}{#1}} 

\usepackage[utf8]{inputenc} 
\usepackage[T1]{fontenc}    
\usepackage{hyperref}       
\usepackage{url}            
\usepackage{booktabs}       
\usepackage{amsfonts}       
\usepackage{nicefrac}       
\usepackage{microtype}      
\usepackage{xcolor}         
\usepackage[linesnumbered,ruled,vlined]{algorithm2e}

\usepackage{amssymb}
\usepackage{latexsym}
\usepackage{amsmath,amssymb}
\usepackage{graphicx}
\usepackage{multirow}
\usepackage{adjustbox}
\usepackage{makecell}
\usepackage[nottoc]{tocbibind}

\usepackage{mathrsfs} 

\begin{document}

\title{Federated Split Vision Transformer for COVID-19 CXR Diagnosis
using Task-Agnostic  Training}

%


\author{
Sangjoon Park$^1$\thanks{Authors contributed equally.} \And Gwanghyun Kim$^1$\footnotemark[\value{footnote}] \AND Jeongsol Kim$^1$ \And Boah Kim$^1$ \And Jong Chul Ye$^{1,2,3}$ \And
  \normalfont  $^1$ Department of Bio and Brain Engineering\\
  $^2$Kim Jaechul Graduate School of AI,
    $^3$Deptartment of Mathematical Sciences\\
   Korea Advanced Institute of Science and Technology (KAIST) \\
  \texttt{\{depecher, gwang.kim, wjdthf3927, boahkim, jong.ye\}@kaist.ac.kr}}

\maketitle

\begin{abstract}
Federated learning, which shares the weights of the neural network across clients, is gaining attention in the healthcare sector as it enables training on a large corpus of decentralized data while maintaining data privacy. 
For example, this enables neural network training for COVID-19 diagnosis on chest X-ray (CXR) images without collecting patient CXR data across multiple hospitals. Unfortunately, the exchange of the weights quickly consumes the network bandwidth if highly expressive network architecture is employed. 
So-called split learning partially solves this problem by dividing a neural network into a client and a server part, so that the client part of the network takes up less extensive computation resources and bandwidth.
However, it is not clear how to find the optimal split without sacrificing the overall network performance. To amalgamate these methods and thereby maximize their distinct strengths, here we show that the Vision Transformer, a recently developed deep learning architecture with straightforward decomposable configuration, is ideally suitable for split learning without sacrificing performance. 
 \add{Even under the non-independent and identically distributed data distribution which emulates a real collaboration between hospitals using CXR datasets from multiple sources}, the proposed framework was able to attain performance comparable to data-centralized training. In addition, the proposed framework along with heterogeneous multi-task clients also improves individual task performances including the diagnosis of COVID-19, eliminating the need for sharing large weights with innumerable parameters. Our results affirm the suitability of Transformer for collaborative learning in medical imaging and pave the way forward for future real-world implementations.

\end{abstract}

\section{Introduction}
After its earlier success in many fields, deep neural networks have found a pervasive suite of applications in healthcare research including medical imaging, becoming a new de facto standard \citep{wang2020deep, gaillochet2020joint, dewey2019deepharmony, tajbakhsh2020embracing, fu2020deep, zhou2021review, chandriah2021maximizing, xu2018less, jing2017automatic}. Training these networks requires a vast amount of data to achieve robust performance \citep{chartrand2017deep, de2018clinically, sun2017revisiting}. Despite the fact that multi-center collaboration is mandatory due to the shortage of labeled data in a single institution, collaboration in healthcare research is heavily impeded by difficulties in data sharing stemming from the privacy issues and limited consent of patients \citep{van2014systematic, rocher2019estimating, wachinger2015brainprint}.

To alleviate this problem, the distributed machine learning methods, devised to enable the computation on multiple clients and servers leaving data to reside on the source devices, can be effectively leveraged for healthcare research \citep{chang2018distributed, sheller2018multi}. Federated learning (FL) is one of these methods which enables model training on a large corpus of decentralized data \citep{konevcny2016federated, mcmahan2017communication, yang2019federated}. However, FL still holds several limitations in that it depends on clients’ computational resources for its client-side parallel computation strategy for update and is not free from privacy concerns \citep{li2020federated, mammen2021federated, vepakomma2018split}. In contrast to FL, another distributed machine learning method, split learning (SL) offers better privacy and requires lower computational resources of clients by splitting the network between clients and the server \citep{gupta2018distributed, vepakomma2018split}, but still possess problems that it shows significant slower convergence than FL and can not learn under non-independent and non-identically distributed (non-IID) data \citep{gao2020end}. 

Especially under unprecedented pandemic of an emerging pathogen like COVID-19, under which direct multi-national collaboration is deterred for prevention of epidemics, the collaboration via these distributed machine learning approaches is becoming increasingly important, since these enable to build a model with performance tantamount to data-centralized learning without any direct sharing of raw data between institutions to offer privacy. 

Recently proposed Vision Transformer (ViT) architecture \citep{dosovitskiy2020image}, inspired by astounding results of Transformer-based models on natural language processing (NLP), have demonstrated impeccable performance on many vision tasks by enabling to model long dependencies within images. Besides this strength, the straightforward design of the Transformer allows to easily decompose the entire network into parts: the head for extracting features from the input image, the Transformer body to model the dependency between features, and the tail used for mapping features to task-specific output. One of the important contributions in this paper is the observation that this configuration is optimal for SL where a network should be split into the parts for clients and servers. In addition, as suggested in \citep{chen2020pre}, the Transformer body with sufficient capacity can be shared between various tasks, being suitable for multi-task learning (MTL) to leverage robust representation from multiple related tasks to enhance the generalization performance of individual tasks.

Accordingly, here we propose a novel Federated Split Task-Agnostic (\textsc{FeSTA}) framework equipped with a Transformer to simultaneously process multiple chest X-ray (CXR) tasks including diagnosis of COVID-19, emulating a real collaboration between several hospitals. To validate the practicability of \textsc{FeSTA}, we also implemented the framework using a friendly federated learning framework (Flower) protocol \citep{beutel2020flower}, confirming seamless integration of various components. 
Experimental results show that our framework can show stable performance even under non-IID settings which is a frequently faced situation for collaboration between hospitals while offering privacy, by amalgamating FL and SL to maximally exploit their main advantages. In addition, we show that the proposed \textsc{FeSTA} Transformer along with MTL improves the performances of individual tasks. In summary, our contributions are two folds:
\begin{itemize}
\item We proposed a novel \textsc{FeSTA} learning framework equipped with the ViT by utilizing its decomposable design to amalgamate the merit of FL and SL.
\item We showed that the model trained with the \textsc{FeSTA} framework can leverage the robust representations from multiple related tasks to improve the performance of the individual task.
\end{itemize}

\section{Related Work}
\paragraph{Distributed machine learning.}

FL is a distributed machine learning approach originally proposed by Google \citep{konevcny2016federated} to enable the training of a model via distributed devices and data. It eliminates the need to aggregate the raw data in a centralized way by enabling the model to be updated on the edge devices (e.g. mobile phones, computers in hospitals).
Specifically, during the training, the server initializes a global model and sends it to each client. The clients then train the model with their local data in parallel and return the updated model to the server. Then, the server aggregates and distributes those models by a method such as federated averaging ($\texttt{FedAvg}$) \citep{mcmahan2017communication} to update the global model. This process (called round) continues repeatedly until the model converges. Though FL enables decentralized training in a privacy-preserving manner, it still holds limitations as it largely depends on the computation resources of clients and is vulnerable to model inversion attacks \citep{wang2019beyond}. 

Different from FL, SL divides the neural network into several sub-networks, and these separated sub-networks are trained under distributed setting \citep{evgeniou2004regularized}. In detail, the first sub-network is trained on the client-side with local data and then passes the feature to the second sub-network located in the server. The server can access the only feature from the first sub-network, and train the second sub-network to send the subsequent feature to the third sub-network on the client. Finally, the third sub-network is able to provide the output of the overall network. By inserting black-box sub-networks in both client and server sides, it is possible to offer better privacy than FL. Besides the privacy benefit, SL uses less computational resources than FL at the client-side. Nevertheless, since the one cycle of forward and backward passes is finished after data and gradients move back and forth across the sub-networks distributed on multiple sides, the convergence of SL is considerably slower than FL. In addition, it was reported that the convergence is not reached at all under the non-IID setting  \citep{gao2020end}.

\paragraph{Multi-task learning.}
MTL is a learning strategy to improve the generalization performance of a specific task with the help of information from other related tasks. In MTL, models for multiple related tasks are trained simultaneously \citep{caruana1998multi, evgeniou2004regularized}. In its early era, the motivation of MTL is to mitigate the data insufficiency problem where the number of labeled data is limited for each task to train an accurate learner. MTL helps to reuse knowledge and thereby reduce the requirement for labeled data for each task, exhibiting that the MTL model can achieve better performance than the single-task counterpart in many fields ranging from computer vision \citep{eigen2015predicting, misra2016cross, rebuffi2017learning, liu2019end} to NLP \citep{luong2015multi, guo2018soft, liu2019multi, ruder2019latent}. 
With increased data, the MTL model can learn more robust representations via knowledge sharing among multiple tasks, resulting in improved performance and less overfitting for the individual task.

\paragraph{Vision Transformer.}
Transformer \citep{vaswani2017attention}, which was originally developed for NLP, is a deep neural network based on an attention mechanism that utilizes an appreciably large receptive field. After achieving state-of-the-art (SOTA) performance in NLP, it has also inspired the vision community to explore its application on vision tasks to utilize its ability to model long-range dependency within an image \citep{khan2021transformers}. The ViT was one of the successful attempts to apply Transformer directly to images, achieving excellent results compared to the SOTA convolutional neural networks in image classification tasks \citep{dosovitskiy2020image}. Furthermore, in addition to the superb performance, the straightforward modular design of ViT facilitates broad applications in many tasks only with minimal change. \add{\citet{chen2020pre} proposed image processing transformer, which is one of the successful multi-task model for various computer vision tasks, by splitting ViT into shared body and task-specific heads and tails, suggesting that Transformer body with sufficient capacity can be shared across relevant tasks. However, they leveraged encoder-decoder design and the usefulness of the multi-task ViT model was not evaluated along with the distributed learning methods.}

Recently, ViT was successfully used for diagnosis and severity prediction of COVID-19, showing the SOTA performance \citep{park2021vision}. Specifically, to alleviate the overfitting problem with limited data available, the overall framework is decomposed into two steps: the pre-trained backbone network to classify common low-level CXR features, which was leveraged in the second step by Transformer for high-level diagnosis and severity prediction of COVID-19. By maximally utilizing the merit of the large-scale database containing more than 220,000 CXR images, the model has attained stable generalization performance as well as SOTA performance in a variety of external test data from different institutions, even with the limited number of labeled cases for COVID-19.

\section{Split Task-Agnostic Transformer for CXR COVID-19 Diagnosis}

Inspired by these works, here we are interested in utilizing ViT for distributed learning in COVID- 19 CXR diagnosis, where the collaboration via these distributed machine learning approaches is becoming increasingly important by offering privacy and still allowing similar performance to the data-centralized learning.

The reason we are interested in ViT architecture for this purpose is that the natural configuration of ViT may be optimal for MTL as well as SL where the easily decomposable modular design of the network is preferred, suggesting a possibility to maximally reconcile the merits of MTL and SL through ViT architecture. {Specifically, the clients just train the head and tail parts of the network, whereas the Transformer body is shared across multiple clients.}
Then, the embedded features from the head network from multiple clients can be leveraged in the second step by Transformer to process individual tasks including diagnosis of COVID-19. Furthermore, by maximally utilizing the merit of the large-scale database from CXR for various tasks, 
{our goal is to demonstrate stable generalization performance as well as SOTA performance in the external test dataset.}


\begin{figure}
    \centering
    \includegraphics[width=0.96\textwidth]{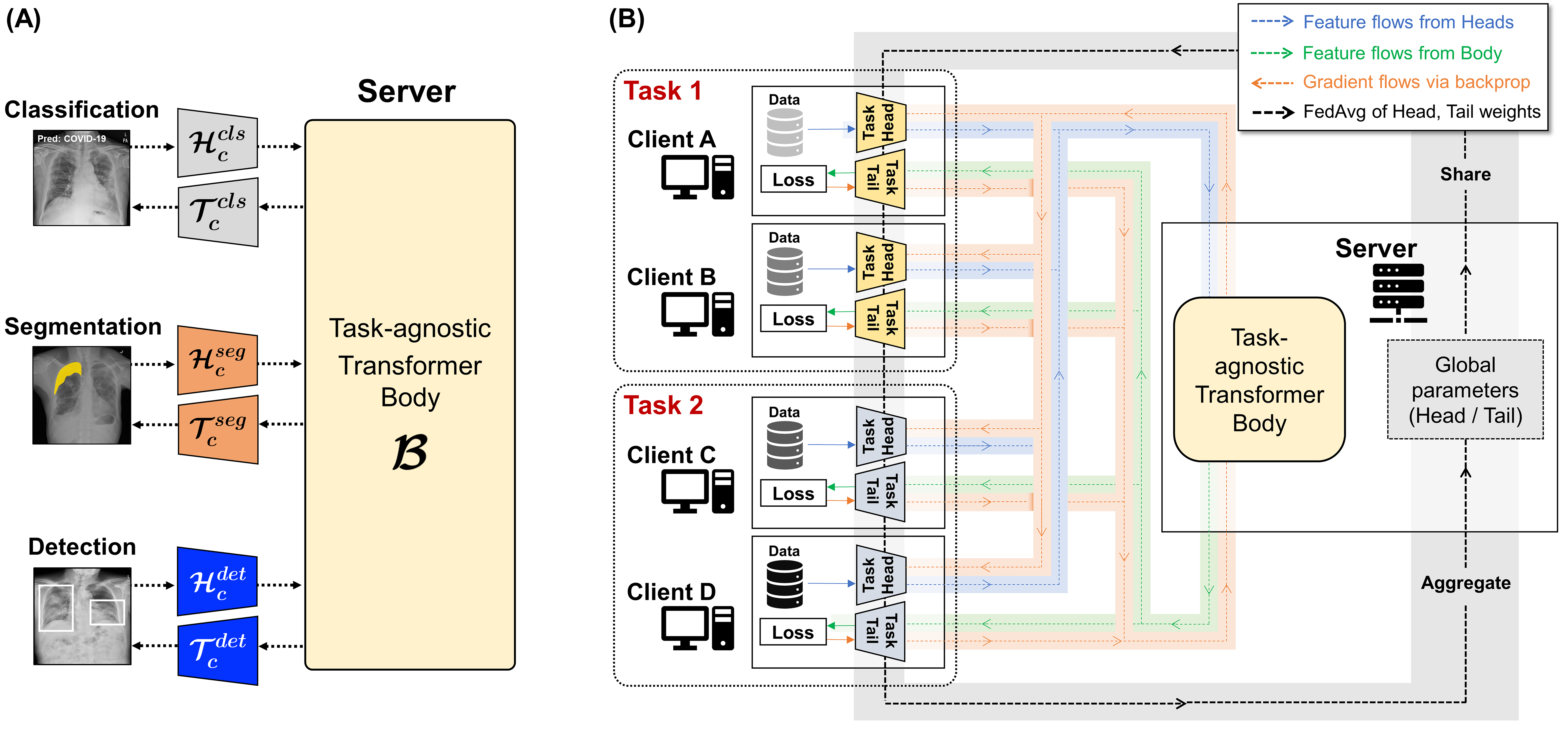}
    \caption{Overall framework of the proposed method. (a) Split task-agnostic CXR Transformer for multi-task learning and (b) the concept of \textsc{FeSTA} learning process.} 
    \label{fig:1}
\end{figure}

\subsection{\textsc{FeSTA}: Federated Split Task-Agnostic Learning}

\add{The concept of the proposed Federated Split Task-Agnostic (\textsc{FeSTA}) learning process is illustrated in Fig~\ref{fig:1}(a) and (b).
Let $\Cc=\bigcup_{k=1}^K C_k $ be a group of client sets with different CXR tasks, where $K$ denotes the number of tasks and $C_k$ has one or more clients with different datasets each other for the $k$-th task, i.e. $C_k=\{c_{1}^{k}, c_{2}^{k}, \dots, c_{N_k}^{k} : N_k \geq 1 \}$. Each client $c \in C_k$ has its own task-specific network architecture for a head $\Hc_c$ and a tail $\Tc_c$,
which are connected to Transformer $\Bc$ in the server. }

In our FeSTA framework, the server first initializes the weights of the Transformer body and task-specific heads, tails for each task $k$. 
Then, it distributes the initialized weights of heads and tails to each client $c \in C_k$. For round {$i = 1,2,\ldots R$}, each client (e.g. hospital) perform the forward propagation on their task-specific head and pass their intermediate feature to the server.
Specifically, using the local training data $\{(x_c^{(i)},y_c^{(i)})\}_{i=1}^{N_c}$, the head network $\Hc_c$ encodes the smashed feature maps $h_c^{(i)}$ and sent it to the server:
$h_c^{(i)} = \Hc_c(x_c^{(i)})$.
 Then, the server-side Transformer body $\Bc$ receives the feature from the all clients and generate the features $b_c^{(i)}$ for $c$ in parallel: $b_c^{(i)} = \Bc(h_c^{(i)})$.
 Resulting smashed features from the Transformer are allocated to task-specific tail to produce the final prediction $\hat y_c^{(i)}=\Tc_c(b_c^{(i)})$ according to each task, and the forward path finishes.
 Subsequently, the loss for each client can be calculated as $\ell_c(y_c^{(i)}, \Tc_c(\Bc(\Hc_c(x_c^{(i)}))))$ where $\ell_c(y,\hat y)$ refers to the $c$ client- specific loss between the target $y$ and the estimate $\hat y$.
 By minimizing the loss with respect to the tail weight, the gradients of the local tails 
 are passed reversely to the server. 
 After receiving the gradients, the server performs the back-propagation on the server-side body model, sends the gradients back to the clients.
 
 Specifically, for the task-agnostic body update, the following optimization problem is solved:
 \begin{align}
   \min_{\Bc} \sum_{c\in \Cc} \sum_{i=1}^{N_c}\ell_c(y_c^{(i)}, \Tc_c(\Bc(\Hc_c(x_c^{(i)})))),
\end{align}
For the task-specific {fine-tuning}, the following optimization problem is solved:
 \begin{align}
   \min_{\Hc_c, \Tc_c} \sum_{i=1}^{N_c}\ell_c(y_c^{(i)}, \Tc_c(\Bc(\Hc_c(x_c^{(i)})))),
\end{align}
 Finally, the server aggregates and averages the weights of local heads and tails from the clients to update the global heads and tails among the clients for the same tasks via $\texttt{FedAvg}$, and distributes back the updated global weights of heads and tails for each task $k$ to the clients. The algorithm is formally presented in Algorithm~\ref{FeSTA}.

 \begin{algorithm}[ht]
    \caption{\textsc{FeSTA}: Federated Split Task-Agnostic learning\label{FeSTA}}
    \DontPrintSemicolon
    \SetKwProg{Fn}{Function}{:}{}
    \SetKwFunction{ServerMain}{ServerMain}
    \SetKwFunction{ClientHead}{ClientHead}
    \SetKwFunction{ClientTail}{ClientTail}
    \SetKwFunction{ClientUpdate}{ClientUpdate}
    \SetKwFor{ForP}{for}{do in parallel}{endfor}
    \SetAlgoNoLine
    \Fn{\ServerMain}{
        \SetAlgoVlined
        Initialize the body weight ${{w}}_\mathcal{B}^{(1)}$ and client head/tail weights $({\bar{w}}_{\Hc, k}, {\bar{w}}_{\Tc, k})$ for each task $k \in \{1,...,K\}$ in server\;
        \For{$\mathbf{rounds} \ i = 1,2,\ldots R$}{
            \ForP{$\mathbf{tasks} \ k \in \{1,2,\ldots K\}$}{
                \ForP{$\mathbf{clients} \ c \in C_k$}{
                    \If{$i = 1 \ \mathbf{or} \ \add{(i-1)} \in \textnormal{UnifyingRounds}$}{
                     Set client $ ({{w}}_{\Hc_c}^{(i)}, {{w}}_{\Tc_c}^{(i)}) \leftarrow ({{\bar{w}}_{\Hc, k}, {\bar{w}}_{\Tc, k}})$  \;
                    }
                    $ h_{c}^{(i)} \leftarrow \ClientHead(c)$ \;
                    $b_{c}^{(i)} \leftarrow \Bc( h_{c}^{(i)})$ \;   
        			${{\partial{L}_{c}^{(i)}}\over  {\partial b_{c}^{(i)}}}  \leftarrow \ClientTail(c,b_{c}^{(i)})$ \& Backprop.\;
                    $({{w}}_{\Hc_c}^{(i+1)}, {{w}}_{\Tc_c}^{(i+1)}) \leftarrow \ClientUpdate(c, {{\partial{L}_{c}^{(i)}}\over {\partial h_{c}^{(i)}} })$ \;
                }
            }
            Update body ${{w}}_\Bc^{(i+1)} \leftarrow {{w}}_{\Bc}^{(i)} - {\eta \over {K}}\sum\limits^{K}_{k=1}\sum\limits_{c \in C_k}{{\partial{L}_{c}^{(i)}}\over N_k{\partial{w}_\Bc^{(i)}} }$ \;
            \If{$i \in \textnormal{UnifyingRounds}$}{
                \For{$\mathbf{tasks} \ k \in \{1,2,\ldots K\}$}{
                     Update $({{\bar{w}}_{\Hc, k}, {\bar{w}}_{\Tc, k}}) \leftarrow ({1 \over N_k}\sum\limits_{c \in C_k}{{w}}_{\Hc_c}^{(i+1)}, \ {1 \over N_k}\sum\limits_{c \in C_k}{{w}}_{\Tc_c}^{(i+1)}) $ \;
                    }
                }
        }
    }
    \Fn{\ClientHead{$c$}}{
        $ x_c \leftarrow$ Current batch of input from client $c$ \;
        \KwRet $\Hc_{c}( x_c)$\;
    }
    \Fn{\ClientTail{$c,b_c$}}{
        $ y_c \leftarrow$ Current batch of label from client $c$\;
        $ L_c \leftarrow {\ell}_c(y_c, \Tc_{c}(b_c)) $ \& Backprop. \;
        \KwRet ${\partial L_c \over  {\partial b_{c}}}$\;
    }
    \Fn{\ClientUpdate{$c, {{\partial{L_c}}\over {\partial h_c} }$}}{
    	Backprop. \& $({{w}}_{\Hc_c}, \ {{w}}_{\Tc_c}) \leftarrow ({{w}}_{\Hc_c} - \eta  {{\partial{L_{c}}}\over {\partial{w}}_{\Hc_c}}, \ {{w}}_{\Tc_c} - \eta {{\partial{L_{c}}}\over {\partial{w}}_{\Tc_c}})$ \;
        \KwRet  $({{w}}_{\Hc_c}, {{w}}_{\Tc_c})$\;
    }
\label{alg:algo1}
\end{algorithm}

 \subsection{Multi-task CXR learning} \label{multi}
 

For synergistic performance improvement, we explore the following three tasks that are commonly used for CXR: classification, segmentation, and object detection. These three tasks were separately trained for the individual model, while used simultaneously to train and to evaluate the task-agnostic model for multiple related tasks. 
The details of each task and dataset are as follows.
\paragraph{COVID-19 classification.} This is the main task we want to achieve through \textsc{FeSTA}.
Since we initialized the weights of classification heads to be the robust feature extractor trained on pre-built large data corpus containing common CXR findings, the classification heads were initialized with the pre-trained weights from CheXpert \citep{irvin2019chexpert} dataset containing 10 CXR findings (no finding, cardiomegaly, opacity, edema, consolidation, pneumonia, atelectasis, pneumothorax, pleural effusion, support device) labeled by experts. \add{From pre-training on the CheXpert dataset, we excluded 32,387 lateral view images, and 29,420 posterior-anterior (PA) and 161,427 anterior-posterior (AP) view data were finally used.} Table~\ref{tab:1} summarizes dataset resources and partitioning for the classification task. We used both public datasets containing labels of infectious disease (Valencian Region Medical Image Bank [BIMCV] \citep{de2020bimcv}, Brixia \citep{BS-Net2021, borghesi2020covid}, National Institutes of Health [NIH] \citep{wang2017chestx}), and CXR data deliberately collected from four hospitals labeled by board-certified radiologists. We put one hospital data aside as an external test dataset to evaluate the classification performance for real-world applications. Overall, 17,183 PA view CXR images were used for training/validation and 365 PA view CXR images for the test. \add{In addition, we performed the experiments in the view-agnostic setting by adding AP view CXRs, to further extend real-world applicability as provided in Appendix C.1. By adding AP view data, the total amounts of CXRs were increased from 17,183 to 24,180 for training/validation and 365 to 556 for external test datasets. Notably, the number of COVID-19 cases was increased from six to 81 in the external test dataset.}
We modeled data from different sources as individual clients, emulating the real-world collaboration between hospitals. This setting is important to validate our method in non-IID data distribution. The study was ethically approved by the Institutional Review Board at each participating hospital and the requirement for informed consent was waived.

\begin{table}[t]
  \centering
  \caption{Datasets and sources for COVID-19 diagnosis}
  \begin{adjustbox}{width=0.93\textwidth}
    \begin{tabular}{lcccccccc}
    \toprule
    \multicolumn{2}{c}{\multirow{3}[5]{*}{\textbf{Total CXR images}}} & \multicolumn{1}{c}{\multirow{2}[3]{*}{\textbf{ External }}} & \multicolumn{6}{c}{\textbf{Training and validation dataset}} \\
\cmidrule{4-9}    \multicolumn{2}{c}{} &       & Client 1 & Client 2 & Client 3 & Client 4 & Client 5 & Client 6 \\
\cmidrule{3-9}    \multicolumn{2}{c}{} & Hospital 1  & Hospital 2   & Hospital 3  & Hospital 4   & NIH   & Brixia & BIMCV \\
    \midrule
    Normal & 13,649 & 320   & 300   & 400   & 8,861 & 3,768 & -     & - \\
    Other infection & 1,468 & 39    & 144   & 308   & 977   & -     & -     & - \\
    COVID-19 & 2,431 & 6     & 8     & 80    & -     & -     & 1,929 & 408 \\
    \midrule
    \textbf{Total CXR} & \textbf{17,548} & \textbf{365}   & \textbf{452}   & \textbf{788}   & \textbf{9,838} & \textbf{3,768} & \textbf{1,929} & \textbf{408} \\
    \bottomrule
    \end{tabular}%
    \end{adjustbox}
    \\
  \label{tab:1}%
\end{table}%

\paragraph{Segmentation.} For the segmentation task, we have used the Society for Imaging Informatics in Medicine and the American College of Radiology Pneumothorax Segmentation Challenge dataset \citep{siim2018pneumothorax} consisting of 12,047 CXR images for training, 3,205 images for testing. The training dataset was divided randomly with a 4:1 ratio into training and validation datasets, and we developed and fine-tuned the model for pneumothorax segmentation with these datasets. Afterward, the segmentation performance of the model was evaluated in testing datasets. Since the data for segmentation was anonymized, it was not possible to divide the dataset according to the sources of acquisition. Instead, we randomly divided the entire dataset into the two clients, emulating the collaboration between two hospitals.

\paragraph{Object detection.} For the object detection task, the model was constructed to detect lung opacities in CXR with the Radiological Society of North America (RSNA) Pneumonia Detection Challenge dataset \citep{rsna2018pneumonia} that consists of CXR images and labels with bounding boxes and detailed class information of 26,684 subjects. We randomly divided the entire dataset with a 3:1 ratio into training and testing datasets, with which model was trained and evaluated. Similar to the segmentation task, as the data contains no information about the sources of acquisition, the dataset was randomly divided into two clients to simulate the collaboration between hospitals.

\section{Experimental Results}


\subsection{Implementation details} \label{imple}
For the head and tail parts of our model, the networks specialized for each task were utilized. For classification, we used the modified version of the network proposed by \citet{ye2020weakly}, which comprises DenseNet combined with Probabilistic Class Activation Map (PCAM) operations for the classification task, since it achieved outstanding performance in the CXR classification competition. For segmentation, we adopt TransUNet \citep{chen2021transunet} tthat inserts a Transformer between the convolutional neural network (CNN) encoder and decoder of UNet \citep{ronneberger2015u} to take advantage of both architectures. Similarly, the 2nd place solution in the RSNA pneumonia detection challenge, which is a modified version of RetinaNet for object detection, was utilized. After splitting these models into head and tail, the feature maps between head and tail were mapped to the feature maps with the same dimension of $16 \times 16 \times 768$, and used as input of Transformer body. The Transformer body, equipped with 12 layers of standard Transformer encoder with 12 attention heads, transforms the feature maps to embed better representation. Then, the resulting feature maps from the body were passed to tails to yield the outcomes.

As suggested in \citet{park2021vision}, the head for classification was first initialized with pre-trained weights from the CheXpert dataset. 
We minimized the cross-entropy loss for the classification task. For the segmentation model, we minimized the binary cross-entropy loss combined with dice and focal loss. Finally, for the detection task, we minimized the sum of box classification, box regression, and image classification losses as suggested by \citet{gabruseva2020deep}.
For all tasks, the batch size was 2 per client, and the warm-up step was 500. We set the number of total rounds to 12,000, and the weights of each clients’ head and tail underwent \texttt{FedAvg} per 100 rounds by the server.
 
\begin{figure}
    \centering
    \includegraphics[width=0.97\textwidth]{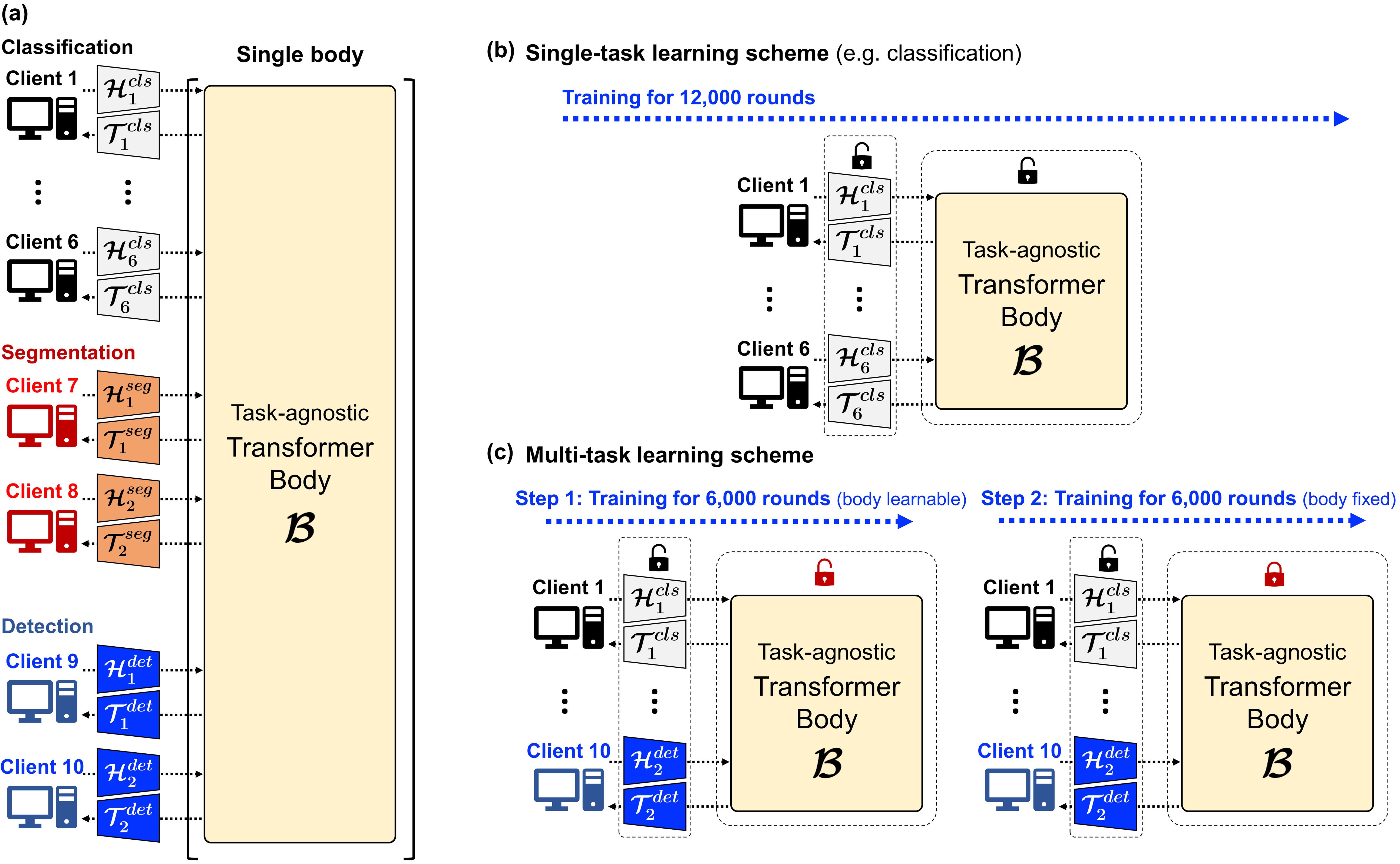}
    \caption{Implementation details of the proposed method. (a) Detailed experimental setting for multi-task learning (MTL) with three different tasks composed of 10 clients (six for classification, two for segmentation, two for detection. Training scheme for (b) single-task learning, and (c) MTL.} 
    \label{fig:2}
\end{figure}

Fig.~\ref{fig:2} illustrates the implementation details and the experimental settings of the proposed method. To simulate single-task learning (STL) in Fig.~\ref{fig:2}(b), we considered each six data sources as different clients for the classification task. 
On the other hand, for MTL shown in Fig.~\ref{fig:2}(c),  10 clients were simultaneously used, with six clients for classification and two clients for segmentation and detection tasks, respectively. For the segmentation and detection tasks, the training data set was randomly split into two subsets, whereas non-IID distribution was used for the classification task. To adjust the loss scale, the customized weights of 1:2:2 were applied for classification, segmentation, and detection tasks to update the common body weights. We divided the MTL into two steps, jointly training the task-specific heads, tails, and the task-agnostic body (6,000 rounds), and fine-tuning only the task-specific heads, tails with the body weights fixed (6,000 rounds). \add{By fixing the parameters of the shared Transformer body during the second step, the best models can be selected for different tasks according to the performance evaluation metrics of each task, even at the different rounds.}

The \textsc{FeSTA}, FL, and SL simulation was performed on the modified version of Flower (licensed under an Apache-2.0 license) \citep{beutel2020flower} FL framework. All experiments were performed with Python version 3.8 and Pytorch version 1.7 on Nvidia RTX 3090, 2080 Ti, and 1080 Ti. For more details of implementation, refer to Appendix A.

\paragraph{Performance metrics.}
We used the area under the receiver operating characteristic curve (AUC) to evaluate the diagnostic performance in the classification task. To evaluate the accuracy of segmentation, the Dice coefficient was used to quantitatively measure the overlap between the segmentation results by model and the ground truths. The detection results were evaluated by calculating the mean average precision (mAP) at the different intersection over union, with a threshold range from 0.4 to 0.75 with a step size of 0.05 as suggested in \citet{rsna2018pneumonia}. All experiments have been run and evaluated with three different random seeds for the weight initialization to prevent the chance of confusing the results.

\subsection{Results}
\paragraph{\textsc{FeSTA} vs. Other strategies.}
We  compared the performance of 
classification model trained with \textsc{FeSTA} with a data-centralized setting and other distributed learning strategies for COVID-19 classification task under the non-IID setting. To simulate data-centralized training, the whole network with connected head, body, and tail was trained with the integrated dataset of all six sources. To simulate FL, the whole network was aggregated and distributed by the server with \texttt{FedAvg} as suggested by \citet{mcmahan2017communication}. On the other hand, for SL, the split sub-networks reside in the clients and server-side, and the client-side sub-networks were not aggregated as in \citep{vepakomma2018split}. The same experimental settings and hyperparameters were applied for a fair comparison. For more details of the data-centralized learning and other distributed learning methods, refer to Appendix B. 

As shown in Table~\ref{tab:tab2}, our method achieved comparable performance to the data-centralized learning method as well as outperformed the existing distributed learning methods, suggesting the superiority of our method over other methods. \add{Of note, the performance could be further enhanced with MTL, which is a distinct strength of the proposed method.}

\add{In our additional experiments after adding AP view data, the model showed even better performance with the increased number of cases (Appendix C.1). This remarkable view-agnostic behavior of the model incentivizes the real-world application of computer-aided diagnosis of COVID-19, as the recent review on artificial intelligence models for COVID-19 diagnosis claims that the AI model, which has been pouring out a lot recently, are not helpful at all from the viewpoint of clinical application \citep{wynants2020prediction}, in which the performances were substantially unstable by the factors like the view of CXR images \citep{roberts2021common}.
}

\begin{table}[h!]
  \centering
  \caption{Comparison for the performance of the proposed method with other strategies}
    \begin{adjustbox}{width=0.85\textwidth}
    \begin{tabular}{lcccc}
    \toprule
    \multicolumn{1}{c}{\multirow{2}[4]{*}{\textbf{Strategy}}} & \multicolumn{4}{c}{\textbf{AUC}} \\
\cmidrule{2-5}          & \textbf{Average} & \textbf{COVID-19} & \textbf{Others} & \textbf{Normal} \\
    \midrule
    Data-centralized & 0.911 $\pm$ 0.016  & 0.883 $\pm$ 0.036  & 0.927 $\pm$ 0.013  & 0.923 $\pm$ 0.004 \\
    Federated learning & 0.891 $\pm$ 0.019  & 0.840 $\pm$ 0.035  & 0.926 $\pm$ 0.018  & 0.906 $\pm$ 0.028 \\
    Split learning & 0.863 $\pm$ 0.005 & 0.807 $\pm$ 0.012 & 0.892 $\pm$ 0.007 & 0.889 $\pm$ 0.019 \\
    \textsc{FeSTA} (single-task learning) & 0.909 $\pm$ 0.021 & 0.880 $\pm$ 0.008 & \ 0.916 $\pm$ 0.038 & 0.931 $\pm$ 0.021 \\
    \add{\textbf{\textsc{FeSTA} (multi-task learning)}} & \add{\textbf{0.931 $\pm$ 0.004}} & \add{\textbf{0.926 $\pm$ 0.023}} & \add{\textbf{0.929 $\pm$ 0.016}} & \add{\textbf{0.938 $\pm$ 0.013}} \\
    \bottomrule
    \end{tabular}%
        \end{adjustbox}
  \label{tab:tab2}%
  \vspace*{0.2cm}\\
  \footnotesize Note: Experiments were performed repeatedly with three random seeds to report mean and standard deviation. For evaluation of split learning, the average metric between clients are calculated.
\end{table}

\paragraph{Multi-task learning vs. Single-task learning.}
We evaluate whether the task-agnostic Transformer body contributes to improving the performance of the entire model by leveraging better representation from the MTL with several related tasks, namely classification, segmentation, and object detection in this work. As provided in Table~\ref{tab:tab3}, the models trained with the MTL approach showed at least comparable or even better performance compared with STL counterparts without the need to create an individual body model for each task or sharing a large body model between clients, which suggest the distinct merit of MTL approach with our framework.

\add{
In addition to the STL models, the MTL models trained with our framework outperformed task-specific expert models and provided comparable or even better performances to Kaggle’s winning solution for the same tasks. Of note, when we substitute the shared layers of the Transformer body with CNN architecture with similar complexity, the performance gain was no longer maintained, suggesting the suitability of Transformer architecture in learning shared representation between the related tasks as provided in the additional experiments in Appendix C.2.
}

\begin{table}[h!]
  \centering
  \caption{Comparison of the performances between single-task and multi-task learning}
      \begin{adjustbox}{width=0.67\textwidth}
    \begin{tabular}{lccccc}
    \toprule
    \multicolumn{1}{c}{\multirow{2}[1]{*}{\textbf{Tasks}}} & \multirow{2}[1]{*}{\textbf{Metrics}} & \multicolumn{2}{c}{\multirow{2}[1]{*}{\textbf{Single-task learning}}} & \multicolumn{2}{c}{\multirow{2}[1]{*}{\textbf{Multi-task learning}}} \\
          &       & \multicolumn{2}{c}{} & \multicolumn{2}{c}{} \\
    \midrule
    \textbf{Classification} & AUC   & \multicolumn{2}{c}{0.909 $\pm$ 0.021} & \multicolumn{2}{c}{\textbf{0.931 $\pm$ 0.004}} \\
    \textbf{Segmentation} & Dice  & \multicolumn{2}{c}{0.798 $\pm$ 0.016} & \multicolumn{2}{c}{\textbf{0.821 $\pm$ 0.003}} \\
    \textbf{Detection} & mAP   & \multicolumn{2}{c}{0.202 $\pm$ 0.008} & \multicolumn{2}{c}{\textbf{0.204 $\pm$ 0.002}} \\
    \bottomrule
    \end{tabular}%
            \end{adjustbox}
  \label{tab:tab3}%
  \\  \vspace*{0.2cm}
  \footnotesize Note: Experiments were performed repeatedly with three random seeds to report mean and standard deviation.
\end{table}%

\paragraph{Model sizes and communicative benefit.}
Table~\ref{tab:tab4} shows the numbers of parameters and the sizes of sub-networks. The task-agnostic body is the largest among sub-networks in terms of both the number of parameters and model size. This suggests that the largest part of the model does not need to be aggregated and distributed between client-client and client-server, which offers substantial communicative benefit. In all tasks, the size of the body is more than half of the entire network. Therefore, not having to share this huge part means the total training time can be reduced considerably. In our experiments, the training time with the proposed method was four times shorter than the FL approach in which the entire network is shared and distributed. In addition, the proposed method offers another benefit of saving the computational resources of the server by processing multiple tasks with a single body model to enable the server to efficiently handle the requests for various tasks from many clients simultaneously.

\add{
Nonetheless, there remains a concern of communicative heavy load between the server and clients caused by the frequent transmission of features and gradients. Our further analysis of communication costs including features and gradients transmission are provided in the additional experiments, which suggests that the communication costs of the proposed \textsc{FeSTA} framework was substantially lower than those of FL, although higher than SL. For detailed results of the computation for communication costs, refer to Appendix C.3.}

    

\begin{table}[h!]
  \centering
  \caption{Parameter numbers and model sizes of the sub-networks}
  \begin{adjustbox}{width=0.87\textwidth}
    \begin{tabular}{lcccccc}
    \toprule
    \multicolumn{1}{c}{\multirow{2}[4]{*}{\textbf{Task}}} & \multicolumn{2}{c}{\textbf{Head}} & \multicolumn{2}{c}{\textbf{Body}} & \multicolumn{2}{c}{\textbf{Tail}} \\
\cmidrule{2-7}          & \textbf{Parameters} & \textbf{Size} & \textbf{Parameters} & \textbf{Size} & \textbf{Parameters} & \textbf{Size} \\
    \midrule
    \textbf{Classification} & 13.313 M & 54.1 MB & \multirow{3}[2]{*}{66.367 M} & \multirow{3}[2]{*}{265.5 MB} & 0.002 M & 11.7 KB \\
    \textbf{Segmentation} & 15.041 M & 60.2 MB &       &       & 7.387 M & 29.6 MB \\
    \textbf{Detection} & 27.085 M & 108.8 MB &       &       & 19.773 M & 79.1 MB \\
    \bottomrule
    \end{tabular}%
    \end{adjustbox}
  \label{tab:tab4}%
  \\ \vspace*{0.2cm}
  \footnotesize Note: Model sizes were estimated by parameter numbers and file sizes of saved weights.
\end{table}

\subsection{Ablation study}
\paragraph{Role of Transformer body.}
\add{As provided in Table~\ref{tab:tab5},} we first performed the ablation study to verify the contribution of the Transformer body on the server. \add{The model without the server-side Transformer body, which is identical to the DenseNet-121 model equipped with PCAM operation \citep{ye2020weakly}, was built} and trained with the same setting, and compared with the proposed model containing the Transformer body. \add{The AUC values were higher with the Transformer body either for STL or MTL compared to those without the Transformer body with the statistical significance, implying the advantageous role of the server-side Transformer body as in Appendix C.4.}

\paragraph{Numbers of round for averaging.} 
Determining the optimal number of rounds for averaging is important. If the aggregation, averaging, and distribution, namely \textsc{FeSTA} is performed too frequently, the cost and required resources for communication increases, which can interfere or even preclude the learning process. On the contrary, if averaged too rarely, naive averaging of the learned model parameters of local learners can be devastating, resulting in nonsensical parameters in some layers \citep{yurochkin2018probabilistic}. Therefore, we performed the ablation to determine the optimal number of rounds to conduct \textsc{FeSTA}. As shown in Table~\ref{tab:tab5}, the averaging per 100 rounds showed the comparable or better performance to less or more frequent counterparts.

More results of ablation studies are provided in Appendix D.

\begin{table}[h!]
  \centering
  \caption{Role of Transformer body and round number for \textsc{FeSTA}}
    
    \begin{adjustbox}{width=0.85\textwidth}
   \begin{tabular}{lcccc}
    \toprule
    \multicolumn{1}{c}{\multirow{2}[4]{*}{\textbf{Method}}} & \multicolumn{4}{c}{\textbf{AUC}}  \\
\cmidrule{2-5}          & \textbf{Average} & \textbf{COVID-19} & \textbf{Others} & \textbf{Normal} \\
    \midrule
    \multicolumn{5}{c}{Role of Transformer body} \\
    \midrule
    w/o Transformer body & 0.889 $\pm 0.015 $ & 0.874 $\pm 0.056$ & 0.895 $\pm 0.011$ & 0.898 $\pm 0.009$$^*$ \\
    \textbf{w Transformer body} & \textbf{0.909 $\pm$ 0.021} & \textbf{0.880 $\pm$ 0.008} & \textbf{0.916 $\pm$ 0.038} & \textbf{0.931 $\pm$ 0.021$^*$} \\
    \midrule
    \multicolumn{5}{c}{Number of rounds for \textsc{FeSTA}} \\
    \midrule
    per 10 rounds & 0.822 $\pm$ 0.023  & 0.724 $\pm$ 0.053  & 0.884 $\pm$ 0.017 & 0.858 $\pm$ 0.035 \\
    \textbf{per 100 rounds} & \textbf{0.909 $\pm$ 0.021} & 0.880 $\pm$ 0.008 & \textbf{0.916 $\pm$ 0.038} & 0.931 $\pm$ 0.021 \\
    per 1000 rounds & 0.903 $\pm$ 0.012 & \textbf{0.905 $\pm$ 0.019} & 0.866 $\pm$ 0.034 & \textbf{0.939 $\pm$ 0.005} \\
    \bottomrule
    \end{tabular}%
    \end{adjustbox}

  \label{tab:tab5}%
  \vspace*{0.2cm}
  \footnotesize Note: Experiments were performed repeatedly with three random seeds to report mean and standard deviation. \\
  \add{Note: $^*$ denotes statistically significant difference.}
\end{table}

\section{Conclusions}
In this paper, we proposed a novel Federated Split Task-Agnostic (\textsc{FeSTA}) framework suitable to leverage the formidable benefit of ViT to simultaneously process multiple CXR tasks including the diagnosis of COVID-19. With the optimal configuration of ViT for modulation, it was possible to surpass the existing methods for distributed learning and achieve the performance comparable to data-centralized learning with our framework, even under the skewed data distribution. Moreover, our framework alongside clients to process multiple related tasks also improves the performances of individual tasks, while eliminating the need to share the data and large weights of the body network. These results suggest the suitability of the Transformer for collaborative learning in medical imaging and pave the way forward for future real-world applications.

\section{Limitation and Potential Negative Societal Impacts} \label{limit}
This work is more like a proof-of-the-concept study, rather than a ready-to-use solution for the industry. Therefore,  it holds some limitations which may occur in a real-world application.  
Recent works on privacy attacks in the FL setting have implied that the belief that "Privacy can be protected by the decentralized nature of the FL" is not true \citep{aono2017privacy, wang2019beyond, zhao2020idlg}. In this work, however, the experimental results regarding the privacy issue such as threatening privacy via inversion attack are not suggested. Secondly, although the unique challenges of FL have been suggested by previous work including the problems of significant bottleneck in communication, stragglers, and fault tolerance which is exacerbated than in typical data-centralized learning \citep{smith2017federated, li2020federated}, we have not conducted the experiments to evaluate the robustness of our method against these unique challenges.

\add{Although distributed learning-enabled learning without sharing data, decentralization is not a panacea against the privacy problem. Similar to the previously distributed learning methods, a potential risk arises from these limitations that our algorithm may not be free from privacy issues via model inversion attack against the server, since the server retains the parameters of the entire network in process of \texttt{FedAvg} of the heads and tails despite the split design of the sub-networks upon both client and server-sides. Although the risk could be mitigated to some degree with the specific settings (e.g. small gradient due to pre-trained backbone, deeper network, more pooling layer, a mixture of multiple tasks), the privacy problem should not be ignored since we aim to use this method in collaboration with hospitals where the patient privacy is a matter of the highest priority. Therefore, the methods to enhance the security such as differential privacy \citep{mcmahan2017learning}, secure multi-party computation \citep{bonawitz2017practical}, data compression \citep{zhu2020deep} and authenticated encryption \citep{rogaway2002authenticated} as well as the recent method to prevent gradient inversion attack without sacrificing FL performance \citep{sun2021soteria} should be utilized along with the proposed method to further reduce the risk of privacy leakage in a real-world application.}

\section*{Acknowledgements and Disclosure of Funding}
\add{This research was funded by the National Research Foundation (NRF) of Korea grant NRF-2020R1A2B5B03001980. This work was also supported by Institute of Information and communications Technology Planning and Evaluation (IITP) grant funded by the Korea government(MSIT) (No.2019-0-00075, Artificial Intelligence Graduate School Program(KAIST)),
and by the KAIST Key Research Institute (Interdisciplinary Research Group) Project.}


\bibliographystyle{abbrvnat}
\bibliography{refs}

\newpage 

\appendix

\section{Implementation Details}\label{appen_A}
Here we describe the details of the hyperparameters used and their implementation. Our experiments were mainly implemented using Python version 3.7.5 and 3.8.8 with packages Pytorch version 1.7.1 and 1.8.1. We used a friendly federated learning framework (Flower) protocol \citep{beutel2020flower} to implement the distributed learning system, Pytorch to implement the neural network.

\subsection{Hyperparameters}
We have searched different hyperparameters for each chest X-ray (CXR) task. For the classification task, we used an SGD optimizer with a warm-up cosine learning rate scheduler with a max learning rate of 0.0005. For the segmentation task, we minimized the binary cross-entropy loss combined with dice and focal loss using Adam \citep{kingma2014adam} with a learning rate of 0.0001 and cosine annealing scheduler with the maximum rounds of 2,000 for single-task and 1,000 for multi-task learning. Finally, the Adam optimizer along with the warm-up constant scheduler was used with the max learning rate of 0.00002. Gradient clipping was implemented, where the max gradient norm of 1.0 was chosen to enhance the rate of convergence, and the batch sizes per client were 2 for all tasks. We experimentally determined the optimal hyperparameters for each task. Table \ref{tab:A2} and Table \ref{tab:A3} provide the detailed hyperparameters used for each task during single-task learning and multi-task learning.

\begin{table}[htbp]
  \centering
  \caption{Hyperparameters used for single-task learning}
  \begin{adjustbox}{width=0.85\textwidth}
    \begin{tabular}{cccc}
    \toprule
    \textbf{ Hyperparameter } & \textbf{ Classification } & \textbf{Segmentation} & \textbf{Detection} \\
    \midrule
    Learning rate (head and tail) & 0.0005 & 0.002 & 0.00002 \\
    Learning rate (body) & 0.0005 & 0.0005 & 0.0005 \\
    Scheduler & warm-up cosine & warm-up cosine annealing & warm-up constant \\
    Number of rounds & 12,000 & 12,000 & 12,000 \\
    Warm-up rounds & 500   & 500   & 500 \\
    Mini-batch size & 2 per client & 2 per client & 2 per client \\
    Maximum number of rounds & -     & 2,000 & - \\
    \bottomrule
    \end{tabular}%
  \label{tab:A2}%
  \end{adjustbox}
\end{table}%

\begin{table}[htbp]
  \centering
  \caption{Hyperparameters used for multi-task learning}
  \begin{adjustbox}{width=0.85\textwidth}
    \begin{tabular}{cccc}
    \toprule
    \textbf{ Hyperparameter } & \textbf{ Classification } & \textbf{Segmentation} & \textbf{Detection} \\
    \midrule
    Learning rate (head and tail) & 0.0005 & 0.002 & 0.00002 \\
    Learning rate (body) & 0.0005 & 0.0005 & 0.0005 \\
    Scheduler & warm-up cosine & warm-up consine annealing & warm-up constant \\
    Number of rounds & 12,000 & 12,000 & 12,000 \\
    Warm-up rounds & 500   & 500   & 500 \\
    Mini-batch size & 2 per client & 2 per client & 2 per client \\
    Maximum number of rounds & -     & 2,000 & - \\
    \bottomrule
    \end{tabular}%
  \label{tab:A3}%
  \end{adjustbox}
\end{table}%

\subsection{Details of network configuration}
\begin{figure}[h]
    \centering
    \includegraphics[width=0.90\textwidth]{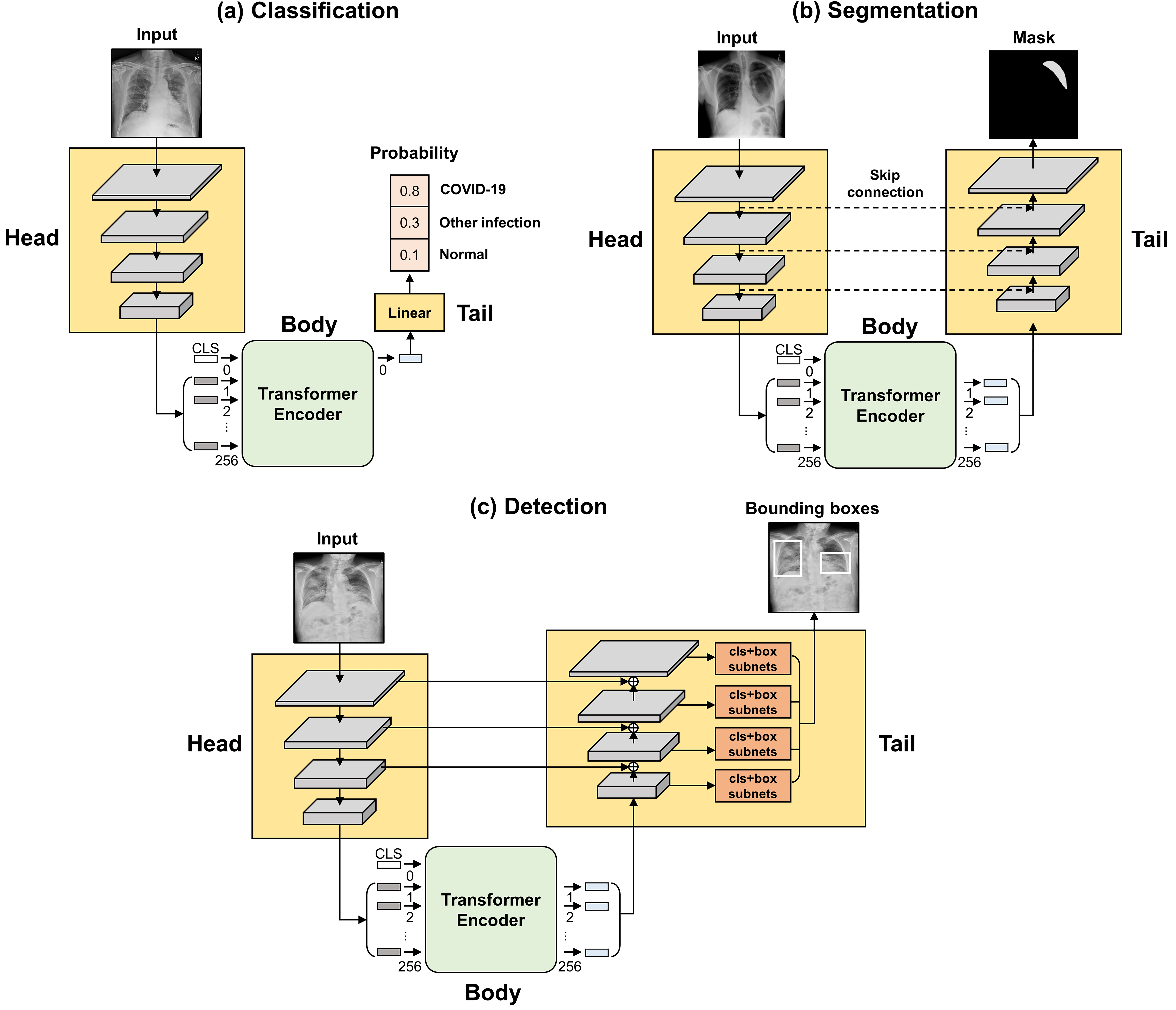}
    \caption{Detailed configuration for (a) classification, (b) segmentation and (c) detection tasks} 
    \label{fig:A1}
\end{figure}

Fig. \ref{fig:A1} depicts the details of network configurations of the proposed method for each task. For classification, the embedded feature of dimension $16 \times 16 \times 768$ from the head is first flattened into the dimension of $256 \times 768$, and used as the input after prepending a \texttt{CLS} token with the same hidden dimension to yield the input of dimension $257 \times 768$. The output from the Transformer body corresponding to this \texttt{CLS} token embed the comprehensive feature of the entire CXR image so that it can be used to make the final prediction (Fig. \ref{fig:A1}(a)). On the other hand, for the segmentation task, the features at the deepest level of TransUNet of the dimension $32 \times 32 \times 1024$ is used as the input of the Transformer after mapping into the dimension of $16 \times 16 \times 768$ and flattened to dimension of $256 \times 768$, and the \texttt{CLS} token is not utilized at all though it is prepended as the same way in the classification task to make the same feature size $257 \times 768$. The resulting transformed features from the body are mapped into original shape and utilized as the same in standard TransUNet architecture (Fig. \ref{fig:A1}(b)). Similarly, the model for the detection task doesn't use the \texttt{CLS} token, and it rather uses a similar approach to that of the segmentation task. The deepest level of the feature pyramid, which has features of the dimension $16 \times 16 \times 1024$, is first mapped into dimension of $16 \times 16 \times 768$, and is used as the input for the Transformer body after flattening and prepending \texttt{CLS} token to make the same dimension of $257 \times 768$ to other tasks. Then, the transformed feature from the body reverts to the original shape and position for the feature pyramid to be combined to yield the final output (Fig. \ref{fig:A1}(c)). 

\subsection{Implementation of our framework upon Flower protocol}
From the implementation perspective of this consequential process, the major hurdle for federated learning (FL) research is the paucity of open source frameworks that support scalable FL on multiple edge devices. Several studies performing FL on millions of edge devices have been published \citep{hard2018federated}, but they are based on a closed industrial system developed by a private corporation and are not publicly available. Meanwhile, even though several open-source frameworks including Tensorflow federated \citep{TensorFl97:online}, PySyft \citep{OpenMine20:online} and LEAF \citep{caldas2018leaf} enabled the experiments on FL simulation, they do not support heterogeneous clients, server-side orchestration and are neither scalable between multiple machines, nor language agnostic. Recently, an open-source framework, Flower has been developed to address this problem which supports the heterogeneous environment and scaling to multiple distributed clients. It offers stable, language- and deep learning framework-agnostic implementation. Moreover, it allows rapid adoption of the existing deep learning algorithm to evaluate their learning dynamics and performances in a federated setting. Therefore, we implemented our framework upon this Flower protocol.

\begin{figure}[h]
    \centering
    \includegraphics[width=1.0\textwidth]{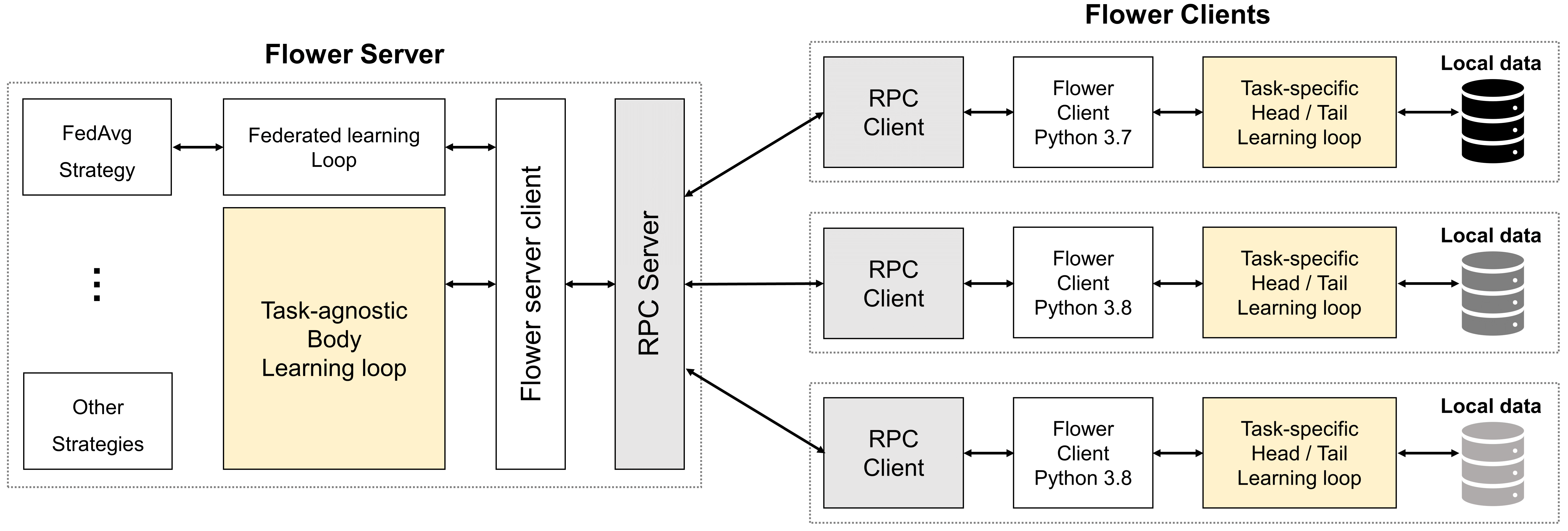}
    \caption{Implementation of our framework on top of Flower protocol.} 
    \label{fig:A2}
\end{figure}

Fig. \ref{fig:A2} illustrates the core components of our framework based on Flower. Since the FL can be considered as an interplay between global (server) and local (client) computations, we implemented the server and client-side components of our framework on top of the Flower server and clients. In Flower clients, task-specific loops of the heads and tails are performed with local data of each client, and the resulting features, gradients, and local parameters are passed toward the RPC client for communication. Then, the remote procedure call (RPC) client communicates with the RPC server in a language-agnostic manner using the bi-directional gRPC stream communication protocol \citep{gRPC38:online}, which offers an efficient binary serialization format. On the server-side, a task-agnostic body loop is performed using the features and gradients received. In addition, the aggregation, distribution of local parameters through a strategy such as federated averaging (\texttt{FedAvg})  are performed per averaging rounds. Finally, the features, gradients, and aggregated global parameters from the server revert to each client.

Different from previous studies that reported the result of single-device simulation \citep{corinzia2019variational}, our method supports the simulation with multiple machines, which is close to real-world implementation of the system across the edge devices.

\section{Data-centralized and Other Distributed Learning Methods}\label{appen_B}

\begin{figure} [h]
    \centering
    \includegraphics[width=1.0\textwidth]{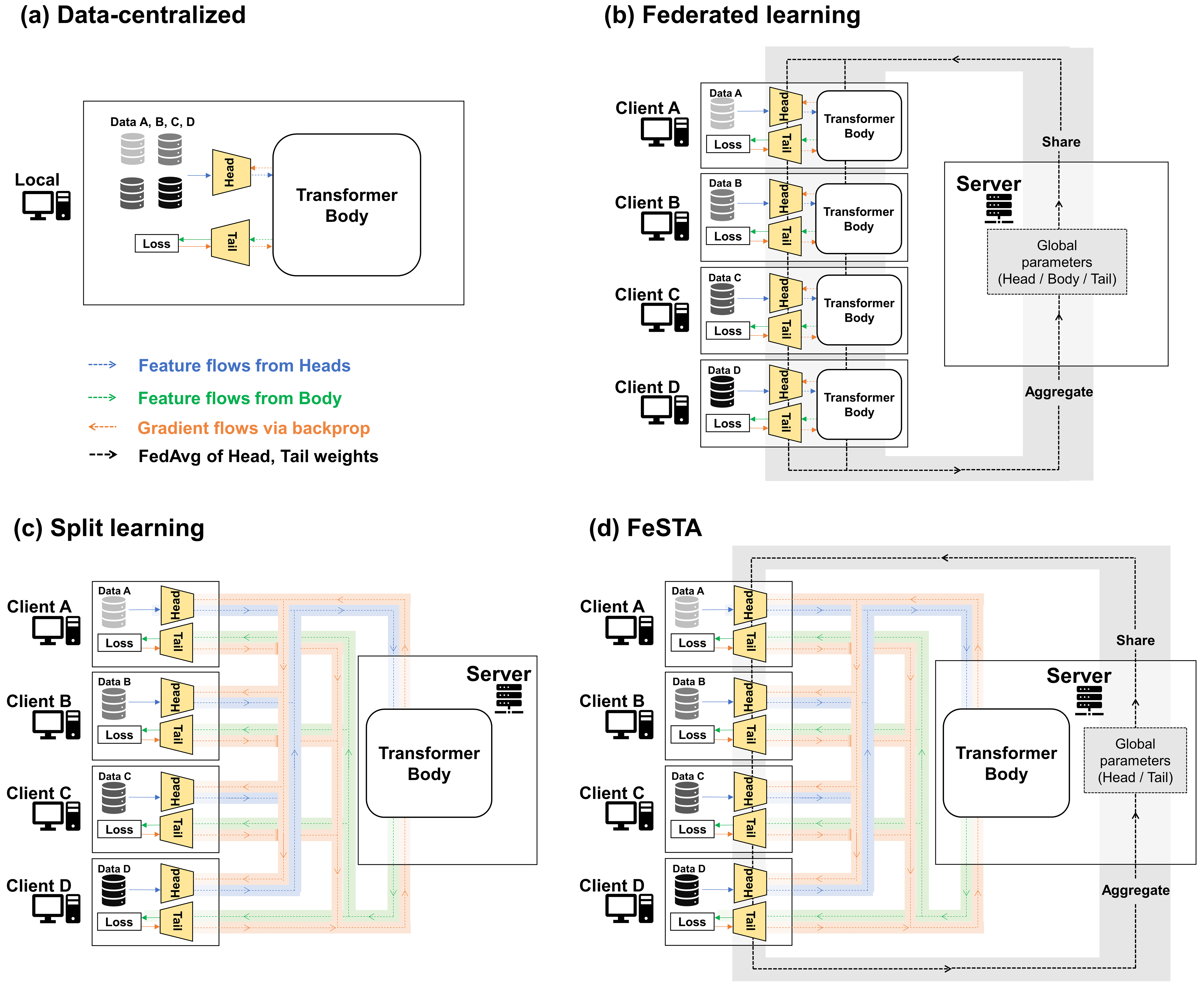}
    \caption{Detailed description for (a) data-centralized learning, (b) federated learning, (c) split learning, (d) \textsc{FeSTA} learning strategies.} 
    \label{fig:A3}
\end{figure}

We perform the comparison of Federated Split Task-Agnostic (\textsc{FeSTA}) with data-centralized and other distributed learning methods on the COVID-19 classification task which is the main task of this study. The details of each learning process are illustrated in Fig.~\ref{fig:A3}.


\subsection{Details of data-centralized learning}
In data-centralized learning, the local data from six clients are centrally aggregated by the server and the single model is trained on a central server as represented in Fig.~\ref{fig:A3} (a). Batch size is set to 12 to match the setting of distributed learning strategies, accounting for batch size two per every six clients. Other settings are used as the same as \textsc{FeSTA} for a fair comparison.

\subsection{Details of federated learning}
In general, the simulation of the FL can be achieved by repeatedly doing three steps, as illustrated: i) update local parameters of the distributed model with local data on each client, ii) send the updated local parameters back to a server for aggregation, iii) distribute the aggregated model back to the clients for next rounds of local updates.   Thus, we trained the entire network consisting of the head, body, and the tail is trained on each client with its local data in parallel without dividing it into sub-networks components as in Fig.~\ref{fig:A3}(b).
This process can be formally written as in Algorithm~\ref{algo:A1}. Regarding the experimental setting, the same settings to those of the proposed \textsc{FeSTA} were used for comparison.

 \begin{algorithm}[ht]
    \caption{Federated learning\label{algo:A1}}
    \DontPrintSemicolon
    \SetKwProg{Fn}{Function}{:}{}
    \SetKwFunction{ServerMain}{ServerMain}
    \SetKwFunction{ClientHead}{ClientHead}
    \SetKwFunction{ClientTail}{ClientTail}
    \SetKwFunction{ClientUpdate}{ClientUpdate}
    \SetKwFor{ForP}{for}{do in parallel}{endfor}
    \SetAlgoNoLine
    \Fn{\ServerMain}{
        \SetAlgoVlined
        Initialize the global weight ${\bar{W}}$ and distribute to each client\;
        \For{$\mathbf{rounds} \ i = 1,2,\ldots R$}{
                \ForP{$\mathbf{clients} \ c \in \Cc$}{
                ${{W}}_{c} \leftarrow \ClientUpdate(c)$ \;
                }
                \If{$i \in \textnormal{UnifyingRounds}$}{
                 Update ${\bar{W}} \leftarrow {1 \over N}\sum\limits_{c \in \Cc}{{W}}_{c} $ \;
                 Distribute the global weight to client $ {{W}}_{c} \leftarrow {\bar{W}}$ for each client $\ c \in \Cc$ \;
                }
            
        }
    }
    \Fn{\ClientUpdate{$c$}}{
        $ x_c, y_c \leftarrow$ Current batch of input \& label from client $c$ \;
        $ L_c \leftarrow {\ell}_c(y_c, \Tc_{c}(\Bc_{c}(\Hc_{c}(x_c)))) $ \& Backprop. \;
        Update ${{W}}_{c} \leftarrow {{W}}_{c} - \eta  {{\partial{L_{c}}}\over {\partial{W}}_{c}}$ \;
        \KwRet  ${{W}}_{c}$\;
    }
\label{alg:algo1}
\end{algorithm}

 \begin{algorithm}[h]
    \caption{Split learning\label{algo:A2}}
    \DontPrintSemicolon
    \SetKwProg{Fn}{Function}{:}{}
    \SetKwFunction{ServerMain}{ServerMain}
    \SetKwFunction{ClientHead}{ClientHead}
    \SetKwFunction{ClientTail}{ClientTail}
    \SetKwFunction{ClientUpdate}{ClientUpdate}
    \SetKwFor{ForP}{for}{do in parallel}{endfor}
    \SetAlgoNoLine
    \Fn{\ServerMain}{
        \SetAlgoVlined
        Initialize the body weight ${{w}}_\mathcal{B}^{(1)}$ and client head/tail weights $({\bar{w}}_{\Hc}, {\bar{w}}_{\Tc})$ in server\;
        \For{$\mathbf{rounds} \ i = 1,2,\ldots R$}{
            \ForP{$\mathbf{clients} \ c \in \Cc$}{
                \If{$i = 1$}{
                 Set client $ ({{w}}_{\Hc_c}^{(i)}, {{w}}_{\Tc_c}^{(i)}) \leftarrow ({{\bar{w}}_{\Hc}, {\bar{w}}_{\Tc}})$  \;
                }
                $ h_{c}^{(i)} \leftarrow \ClientHead(c)$ \;
                $b_{c}^{(i)} \leftarrow \Bc( h_{c}^{(i)})$ \;   
    			${{\partial{L}_{c}^{(i)}}\over  {\partial b_{c}^{(i)}}}  \leftarrow \ClientTail(c,b_{c}^{(i)})$ \& Backprop.\;
                $\ClientUpdate(c, {{\partial{L}_{c}^{(i)}}\over {\partial h_{c}^{(i)}} })$ \;
            }
            Update body ${{w}}_\Bc^{(i+1)} \leftarrow {{w}}_{\Bc}^{(i)} - {\eta \over N}\sum\limits_{c \in \Cc}{{\partial{L}_{c}^{(i)}}\over {\partial{w}_\Bc^{(i)}} }$ \;
        }
    }
    \Fn{\ClientHead{$c$}}{
        $ x_c \leftarrow$ Current batch of input from client $c$ \;
        \KwRet $\Hc_{c}( x_c)$\;
    }
    \Fn{\ClientTail{$c,b_c$}}{
        $ y_c \leftarrow$ Current batch of label from client $c$\;
        $ L_c \leftarrow {\ell}_c(y_c, \Tc_{c}(b_c)) $ \& Backprop. \;
        \KwRet ${\partial L_c \over  {\partial b_{c}}}$\;
    }
    \Fn{\ClientUpdate{$c, {{\partial{L_c}}\over {\partial h_c} }$}}{
    	Backprop. \& $({{w}}_{\Hc_c}, \ {{w}}_{\Tc_c}) \leftarrow ({{w}}_{\Hc_c} - \eta  {{\partial{L_{c}}}\over {\partial{w}}_{\Hc_c}}, \ {{w}}_{\Tc_c} - \eta {{\partial{L_{c}}}\over {\partial{w}}_{\Tc_c}})$ \;
    }
\label{alg:algo1}
\end{algorithm}

\subsection{Details of split learning}

To simulate split learning (SL), we adopted the SL without label sharing as suggested in the original paper of SL \citep{vepakomma2018split}. The detailed process of the SL method used in our experiment can be presented as in Algorithm~\ref{algo:A2}. The overall process of SL is similar to \textsc{FeSTA} except for the fact that a step of aggregation and distribution by the central server is absent in SL as in Fig.~\ref{fig:A3}(c). The splitting configuration of head, body, and tail on client and server sides were the same as in the proposed \textsc{FeSTA}. Since the local head and tail parameters of individual clients are not unified in SL, the inference results on the external testing dataset can be different between clients. Therefore, we calculated evaluation metrics for every six clients and averaged them to get the final score. The other experiment settings, including batch size and learning rate, remain the same as in the proposed \textsc{FeSTA}.

\section{\add{Additional Experiments}}\label{appen_D}
\add{In this section, the results of additional experiments to further analyze the proposed \textsc{FeSTA} learning method are suggested.}

\subsection{\add{Performances with increased number of COVID-19 cases}}
\add{To provide more robust results using the larger corpus of data especially in terms of the number of COVID-19 cases, additional experiments were performed as follows.}

\add{We first swapped the hospital 1 data (containing 6 COVID-19 cases), which was originally used as the external test dataset, with the hospital 3 data (containing 80 COVID-19 cases), and repeated the experiments with the same setting. As suggested in Table~\ref{tab:A7}, the proposed model retained stable performance in hospital 3 data with 80 COVID-19 cases.}

\begin{table}[h!]
  \centering
  \caption{Number of COVID-19 cases with the different external set and the classification performances (AUC).}
    \begin{adjustbox}{width=0.96\textwidth}
    \begin{tabular}{cccccc}
    \toprule
    \textbf{External test set} & \textbf{COVID-19 cases} & \textbf{Average} & \textbf{COVID-19} & \textbf{Others} & \textbf{Normal} \\
    \midrule
    Hospital 1 & 6     & 0.909 ± 0.021 & 0.880 ± 0.008 & 0.916 ± 0.038 & 0.931 ± 0.021 \\
    Hospital 3 & 80    & 0.913 ± 0.019 & 0.871 ± 0.043 & 0.932 ± 0.007 & 0.935 ± 0.015 \\
    \bottomrule
    \end{tabular}%
    \end{adjustbox}
  \label{tab:A7}%
  
\vspace*{0.2cm}
  \footnotesize Note: Experiments were performed repeatedly with three random seeds to report mean and standard deviation.
\end{table}%

\add{Secondly, the additional analysis was performed by holding out all four hospital data (private) from the training set and by using them for external validation. Here, the label system had to be simplified into two categories, COVID-19 and non-COVID-19, as public datasets do not contain any label data for the "other infection" class. The proposed model presented stable performance even after excluding all private CXR data from the training set, dispelling the worry of data leakage problems. Although the performances were slightly decreased, it should be taken into consideration that the total amount of training data decreased to less than half of the original training data by removing the hospital 4 data (Table~\ref{tab:A8}).
}

\begin{table}[h!]
  \centering
  \caption{Classification performances (AUC) of the proposed model using all four hospital datasets as an external testset.}
    \begin{adjustbox}{width=0.64\textwidth}
    \begin{tabular}{ccc}
    \toprule
    \textbf{External test set} & \textbf{COVID-19 cases} & \textbf{COVID-19} \\
    \midrule
    All four hospitals (hospital 1-4) & 94    & 0.879 ± 0.043 \\
    \bottomrule
    \end{tabular}%
    \end{adjustbox}
  \label{tab:A8}%
  
\vspace*{0.2cm}
  \footnotesize Note: Experiments were performed repeatedly with three random seeds to report mean and standard deviation.
\end{table}%

\add{Finally, we gathered additional anterior-posterior (AP) view CXR data labeled by the experts and combined them with the original posterior-anterior (PA) view data as shown in Table~\ref{tab:A9}. The total amount of COVID-19 data has doubled, and COVID-19 cases in hospital 1 increased 6 to 81 CXRs. When adding the additional AP view CXRs and evaluating the performances in hospital 1 data, the performances of the proposed model were not compromised and rather increased especially for the diagnosis of COVID-19 as in Table~\ref{tab:A10}. 
}

\begin{table}[h!]
  \centering
  \caption{Increased dataset and sources for COVID-19 diagnosis.}
    \begin{adjustbox}{width=0.96\textwidth}
    \begin{tabular}{ccccccccc}
    \toprule
    \textbf{CXR view} & \textbf{Total} & \textbf{Hospital 1} & \textbf{Hospital 2} & \textbf{Hospital 3} & \textbf{Hospital 4} & \textbf{NIH} & \textbf{Brixia} & \textbf{BIMCV} \\
    \midrule
    \textbf{AP view (added)} &       &       &       &       &       &       &       &  \\
    Normal & 3662  & 97    & -     & -     & 117   & 3355  & -     & 93 \\
    Other infection & 204   & 19    & 76    & 92    & 17    & -     & -     & - \\
    COVID-19 & 3322  & 75    & 278   & 213   & -     & -     & 2384  & 372 \\
    Total AP CXRs & 7188  & 191   & 354   & 305   & 134   & 3355  & 2384  & 465 \\
    \\
    \textbf{All view (total)} &       &       &       &       &       &       &       &  \\
    Normal & 17311 & 417   & 300   & 400   & 8978  & 7123  & -     & 93 \\
    Other infection & 1672  & 58    & 220   & 400   & 994   & -     & -     & - \\
    COVID-19 & 5753  & 81    & 286   & 293   & -     & -     & 4313  & 780 \\
    Total CXRs & 24736 & 556   & 806   & 1093  & 9972  & 7123  & 4313  & 873 \\
    \bottomrule
    \end{tabular}%
    \end{adjustbox}
  \label{tab:A9}%
  
\end{table}%

\begin{table}[h!]
  \centering
  \caption{Number of COVID-19 cases after adding AP view CXRs and the classification performances (AUC).}
    \begin{adjustbox}{width=1.0\textwidth}
    \begin{tabular}{cccccc}
    \toprule
    \textbf{External test set} & \textbf{COVID-19 cases} & \textbf{Average} & \textbf{COVID-19} & \textbf{Others} & \textbf{Normal} \\
    \midrule
    Hospital 1 (PA data) & 6     & 0.909 ± 0.021 & 0.880 ± 0.008 & 0.916 ± 0.038 & 0.931 ± 0.021 \\
    Hospital 1 (PA and AP data) & 81    & 0.924 ± 0.006 & 0.943 ± 0.015 & 0.879 ± 0.007 & 0.949 ± 0.008 \\
    \bottomrule
    \end{tabular}%
    \end{adjustbox}
  \label{tab:A10}%
  
\vspace*{0.2cm}
  \footnotesize Note: Experiments were performed repeatedly with three random seeds to report mean and standard deviation.
\end{table}%

\subsection{\add{Comparison with task-specific expert and CNN-based multi-task learning models}}
\add{Table~\ref{tab:A5} shows a comparison of the performances of each task between the proposed Transformer-based multi-task learning model trained with \textsc{FeSTA} method and others. First, we compared the proposed MTL model with single task experts, defined as following for each task.}

\begin{itemize}
    \item \textbf{Classification:} DenseNet-121 (D121) model with Probabilistic Class Activation Map operation \citet{ye2020weakly}
    \item \textbf{Segmentation:} AlbuNet \citep{shvets2018automatic} based segmentation network (1st place model in Kaggle SIIM-ACR pneumothorax segmentation challenge \citep{siim2018pneumothorax})
    \item \textbf{Detection:} RetinaNet \citep{lin2017focal} model with SE-ResNext-50 encoder (2nd place model in Kaggle RSNA pneumonia detection challenge \citep{rsna2018pneumonia})
\end{itemize}

\add{As provided in Table~\ref{tab:A5}, the proposed MTL model outperformed the task-specific experts for each specific task. Of note, when the shared Transformer body was substituted with the shared convolutional neural network (CNN) layer for MTL, the performance was substantially dropped in the detection task. Combined together, the results demonstrated the value of the Transformer architecture leveraging global attention as well as local attention, which is suitable for MTL and cannot be substituted by other architecture like shared CNN layers.}

\begin{table}[h!]
  \centering
  \caption{Comparison of performances with task-specific experts and CNN-based MTL models}
    \begin{adjustbox}{width=0.92\textwidth}
    \begin{tabular}{ccccc}
    \toprule
    \textbf{Tasks} & \textbf{Metrics} & \textbf{Task-specific experts} & \textbf{CNN-based MTL} & \textbf{Transformer-based MTL} \\
    \midrule
    \textbf{Classification} & AUC   & 0.898 ± 0.004 & 0.907 ± 0.011 & \textbf{0.931 ± 0.004} \\
    \textbf{Segmentation} & Dice  & 0.736 ± 0.014 & 0.797 ± 0.018 & \textbf{0.821 ± 0.003} \\
    \textbf{Detection} & mAP   & 0.190 ± 0.006 & 0.159 ± 0.035 & \textbf{0.204 ± 0.002} \\
    \bottomrule
    \end{tabular}%
    \end{adjustbox}
  \label{tab:A5}%
  
  \vspace*{0.2cm}
  \footnotesize Note: Experiments were performed repeatedly with three random seeds to report mean and standard deviation.
\end{table}%

\add{In addition, when compared with Kaggle's winning solutions available for the segmentation \citep{siim2018pneumothorax} and detection tasks \citep{rsna2018pneumonia}, the proposed MTL model showed comparable performances as shown in Table~\ref{tab:A6}, suggesting that the Transformer body do not deface the performances of the individual tasks.}

\begin{table}[h!]
  \centering
  \caption{Comparison with Kaggle winning solutions for segmentation and detection tasks}
    \begin{adjustbox}{width=0.96\textwidth}
    \begin{tabular}{cccc}
    \toprule
    \textbf{Segmentation} & \textbf{Dice} & \textbf{Detection} & \textbf{mAP} \\
    \midrule
    1st place solution (description) & 0.764 ± 0.007 & 2nd place solution (SE-ResNext-50) & \textbf{0.211 ± 0.003} \\
    4th place solution (descrption) & \textbf{0.841 ± 0.004} & 2nd place solution (SE-ResNext-101) & 0.199 ± 0.003 \\
    Proposed MTL model  & 0.821 ± 0.003 & Proposed MTL model  & 0.204 ± 0.002 \\
    \bottomrule
    \end{tabular}%
    \end{adjustbox}
  \label{tab:A6}%
  
\vspace*{0.2cm}
  \footnotesize Note: Experiments were performed repeatedly with three random seeds to report mean and standard deviation.
\end{table}%

\subsection{\add{Estimates of communication costs}}
\add{With the intrinsic property of \textsc{FeSTA} learning, a high computational burden is imposed on the server-side device, and this configuration is what we intended. Suppose, if most of the computation is performed on client-sides, all participating hospitals should have devices with high computational capacity. Forcing to prepare high computational resources for participants will obviously hinder the widespread adoption in a real-world application, and preparing a powerful server-side device with better security is rather practical. Nevertheless, there still remains a problem of computational costs between the server and clients. The communication costs between the server as the client can be estimated as follow.}

\add{When the period between averaging is $k$ and transmission of features, gradients, and network parameters are $F$, $G$ and $P$ respectively, total transmission from Server to Client $T$ can be represented as follows:}
 \begin{align}
   T=  k\times(F + G)  + P ,
\end{align}
\add{When parameter numbers of head, body, and tail are $P_h$, $P_b$ and $P_t$ respectively and $k$ is 100, $T$ for each distributed learning strategy can be formulated as follows:}
 \begin{gather}
   T_\text{FL} =  P_h + P_b + P_t , \\
   T_\text{SL} = 100 \times (F + G) , \\
   T_\text{\textsc{FeSTA}} = 100\times(F + G) + (P_h + P_t),
\end{gather}

\add{If the transmission from Server to Client $T$ and that from Client to Server $T_{\text{C} \rightarrow \text{S}}$ are assumed to be equal ($T_{\text{C} \rightarrow \text{S}} = T$), total transmission $T'$ is as follows:}
 \begin{align}
   T'=2T.
\end{align}

 \add{We then calculated the communication costs for feature/gradient transmission and parameter transmission per 1 averaging (=100 rounds) for each task as shown in Table~\ref{tab:A13}. Despite the fact that the communication cost of the proposed  \textsc{FeSTA} framework was larger than that of SL, it was substantially lower than that of FL.}

\begin{table}[h!]
  \centering
  \caption{Communication costs of the distributed learning methods during training per 1 averaging (=100 rounds)}
    \begin{adjustbox}{width=0.75\textwidth}
   \begin{tabular}{lccc}
    \toprule
                 & \textbf{\shortstack[c]{Total \\ transmission}} & \textbf{\shortstack[c]{Feature and gradient \\ transmission}} & \textbf{\shortstack[c]{Network parameter \\ transmission}} \\
    \midrule
        \textbf{Classification} &  &  &  \\ 
        \midrule
        Federated learning & 159.365M & - & 159.365M \\
        Split learning & 78.950M & 78.950M & - \\ 
        \textsc{FeSTA} & 105.580M & 78.950M & 26.630M \\ 
        \midrule
        \textbf{Segmentation} &  &  &  \\
        \midrule
        Federated learning & 177.592M & - & 177.592M \\ 
        Split learning & 78.950M & 78.950M & - \\ 
        \textsc{FeSTA} & 123.808M & 78.950M & 44.858M \\ 
        \midrule
        \textbf{Detection} &  &  &  \\ 
        \midrule
        Federated learning & 226.450M & - & 226.450M \\ 
        Split learning & 78.950M & 78.950M & - \\ 
        \textsc{FeSTA} & 172.665M & 78.950M & 93.715M \\ 
    
    \bottomrule
    \end{tabular}%
    \end{adjustbox}
  \label{tab:A13}%
  \vspace*{0.2cm}
\end{table}

\subsection{\add{Statistical analysis}}
\add{We also performed whether or not the performance gains with the Transformer architecture are statistically significant. As provided in Table~\ref{tab:A11}, the proposed MTL model outperformed the model without the Transformer body with statistical significance, and the performance increase was more prominent in the MTL model.}

\begin{table}[h!]
  \centering
  \caption{Statistical comparison of performances between the model with and without the transformer.}
    \begin{adjustbox}{width=1.0\textwidth}
    \begin{tabular}{ccccccc}
    \toprule
          & \multicolumn{2}{c}{\textbf{COVID-19}} & \multicolumn{2}{c}{\textbf{Others}} & \multicolumn{2}{c}{\textbf{Normal}} \\
\cmidrule{2-7}          & \textbf{AUC (95\% CI)} & \textbf{p-value} & \textbf{AUC (95\% CI)} & \textbf{p-value} & \textbf{AUC (95\% CI)} & \textbf{p-value} \\
    \midrule
    \textbf{w/o Transformer} & 0.867 (0.696 - 1.000) & -     & 0.883 (0.817 - 0.948) & -     & 0.889 (0.837 - 0.941) & - \\
    \textbf{w Transformer (STL)} & 0.868 (0.749 - 0.987) & 0.988 & \textbf{0.905 (0.852 - 0.958)} & 0.498 & 0.927 (0.889 - 0.965) & \textbf{0.019} \\
    \textbf{w Transformer (MTL)} & \textbf{0.945 (0.896 - 0.995)} & 0.266 & 0.893 (0.833 - 0.954) & 0.768 & \textbf{0.938 (0.903 - 0.974)} & \textbf{0.010} \\
    \bottomrule
    \end{tabular}%
    \end{adjustbox}
  \label{tab:A11}%
  
\vspace*{0.2cm}
  \footnotesize Note: For statistical comparison, p-values and Confidence Intervals (CIs) were calculated using DeLong's test. \\
  Note: To evaluate the statistical significance, the models with medium performance were compared.
\end{table}%

\section{Additional Ablation Studies in Multi-Task Learning Setting.}\label{appen_C}
Additional ablation studies have been performed to examine the effect of Transformer body capacity and different training schemes,
as shown in Table \ref{tab:A4}.

\paragraph{Effect of Transformer body capacity}
We first evaluated the effect of the network capacity of the task-agnostic body model on the performances of individual tasks. Since the Transformer processes the task-agnostic modeling between features in a multi-task setting, there exists a possibility that the performance can further increase with the use of a dedicated server system with higher computational resources, once the model shows the performance proportional to the capacity of the Transformer body. As suggested in Table ~\ref{tab:A4}, the model equipped with a smaller body showed lower performance than that of a standard Transformer body equipped with 12 heads and 12 layers, suggesting the possibility of further improvement in performance with a Transformer with higher capacity.

\paragraph{Training scheme}
Since our framework consists of the server-side and client-side sub-networks, it is possible to train only part of these sub-networks or train this sub-network after fixing the others. Thus, we experimented with various training schemes to evaluate their effect on the performance in multi-task settings. For the one-step learning approach, we trained the model after having all sub-networks, namely head, body, and tail, learnable for the entire training round. For the alternating approach, we alternately fixed and unfixed the parameters of the body and head/tail per 100 rounds. As shown in Table~\ref{tab:A4}, both of these approaches show lower performance than the proposed two-step learning approach. This suggests that the simultaneous or alternating approach to train these sub- network components makes training unstable. Fixing the body for multi-task processing after certain rounds and fine-tuning the task-specific components may help to reach the better local minimum for the head and tail for each task, resulting in better generalization performance.

\begin{table}[h!]
  \centering
  \caption{Additional ablations in multi-task setting}
    
    \begin{adjustbox}{width=0.83\textwidth}
     \begin{tabular}{lrrr}
    \toprule
    \multicolumn{1}{c}{\multirow{2}[4]{*}{\textbf{Tasks}}} & \multicolumn{1}{c}{\textbf{Classification}} & \multicolumn{1}{c}{\textbf{Segmentation}} & \multicolumn{1}{c}{\textbf{Detection}} \\
\cmidrule{2-4}          & \multicolumn{1}{c}{\textbf{AUC}} & \multicolumn{1}{c}{\textbf{Dice}} & \multicolumn{1}{c}{\textbf{mAP}} \\
    \midrule
    \multicolumn{4}{c}{Effect of Transformer body capacity} \\
    \midrule
    $H = 4, L = 4, D_{hidden} = 256$ & \multicolumn{1}{l}{0.916 $\pm$ 0.011} & \multicolumn{1}{l}{0.809 + 0.030} & \multicolumn{1}{l}{0.200 $\pm$ 0.007} \\
    $H = 8, L = 8, D_{hidden} = 512$ & \multicolumn{1}{l}{0.916 $\pm$ 0.013} & \multicolumn{1}{l}{0.826 + 0.001} & \multicolumn{1}{l}{0.191 $\pm$ 0.016} \\
    $\boldsymbol{H = 12, L = 12, D_{hidden} = 768}$ &  \textbf{0.931 $\pm$ 0.004} & \textbf{0.821 $\pm$ 0.003} & \textbf{0.204 $\pm$ 0.002}\\
    \midrule
    \multicolumn{4}{c}{Effect of training strategy} \\
    \midrule
    One-step approach & \multicolumn{1}{l}{0.930 $\pm$ 0.022} & \multicolumn{1}{l}{0.801 $\pm$ 0.024} & \multicolumn{1}{l}{0.188 $\pm$ 0.020} \\
    Alternating approach & \multicolumn{1}{l}{0.915 $\pm$ 0.011} & \multicolumn{1}{l}{0.799 $\pm$ 0.021} & \multicolumn{1}{l}{0.179 $\pm$ 0.003} \\
    \textbf{Two-step approach} & \textbf{0.931 $\pm$ 0.004} & \textbf{0.821 $\pm$ 0.003} & \textbf{0.204 $\pm$ 0.002} \\
    \bottomrule
    \end{tabular}%
    \end{adjustbox}

  \label{tab:A4}%
  \vspace*{0.2cm}
  \footnotesize Note: Experiments were performed repeatedly with three random seeds to report mean and standard deviation.
\end{table}

\section{\add{Details of Hospital Dataset}}
\add{Table~\ref{tab:A12} describes the details about the CXR and clinical characteristics of four hospital data deliberately collected for this study.}

\begin{table}[h!]
  \centering
  \caption{Details of CXR and patient characteristics of hospital datasets.}
  \begin{adjustbox}{width=0.84\textwidth}
    \begin{tabular}{ccc}
    \toprule
    \textbf{Data} & \textbf{              Hospital 1} & \textbf{        Hospital 2} \\
    \midrule
    \textbf{CXR image details} &       &  \\
    \textbf{Number of CXRs} & 365   & 452 \\
    \textbf{Modality} & CR (93.7\%), N/A (6.3\%) & CR (99.8\%), N/A (0.2\%) \\
    \textbf{Exposure time (msec)} & 6.7 ± 3.4 & 16.5 ± 7.7 \\
    \textbf{Tube current (mA)} & 473.3 ± 198.1 & 307.8 ± 36.4 \\
    \textbf{Bits} & 12 (12-14) & 12 (12-12) \\
    \textbf{Clinical details} &       &  \\
    \textbf{Age} & 45.8 ± 15.9 & 50.9 ± 17.7 \\
    \textbf{Sex} & M (47.7\%), F (45.2\%), N/A (7.1\%) & M (50.2\%), F (48.8\%) \\
    \textbf{COVID-19 severity} & 1 (1-3) & 5.5 (1-6) \\
    \textbf{CT positive cases} & N/A (100\%) & N/A (100\%) \\
    \textbf{Country} & South Korea & South Korea \\
    \midrule
          &       &  \\
    \midrule
    \textbf{Data} & \textbf{                Hospital 3} & \textbf{              Hospital 4} \\
    \midrule
    \textbf{CXR image details} &       &  \\
    \textbf{Number of CXRs} & 788   & 9838 \\
    \textbf{Modality} &       CR (100\%)         & CR (3.8\%), DX (96.2\%) \\
    \textbf{Exposure time (msec)} & 11.6 ± 8.0 & 8.9 ± 3.9 \\
    \textbf{Tube current (mA)} & 317.8 ± 30.7 & 298.9 ± 43.6 \\
    \textbf{Bits} & 12 (12-14) & 14 (10-15) \\
    \textbf{Clinical details} &       &  \\
    \textbf{Age} & 46.9 ± 16.6 & 46.3 ± 14.4 \\
    \textbf{Sex} & M (26.9\%), F (34.0\%), N/A (39.1\%) & M (49.1\%), F (47.0\%), N/A (3.9\%) \\
    \textbf{COVID-19 severity} & 5 (1-6) & - \\
    \textbf{CT positive cases} & Positive (3.8\%), N/A (96.3\%) & - \\
    \textbf{Country} & South Korea & South Korea \\
    \bottomrule
    \end{tabular}%
    \end{adjustbox}
  \label{tab:A12}%
  
\vspace*{0.2cm}
  \footnotesize Abbreviations: CXR, chest X-ray; CR, computed radiography; DX, digital x-ray; M, male; F, female. \\
  \footnotesize Note: Values are presented as mean ± standard deviation or median (range).
  
\end{table}%

\section{\add{Ethic Committee Approval and Permission Information.}}\label{appen_E}
\paragraph{\add{Ethic Committee Approval and Permission Information}}
\add{The four hospital data deliberately collected for COVID-19 classification were ethically approved by the Institutional Review Board of each hospital. According to their terms of use, the public datasets for the classification task (CheXpert \citep{irvin2019chexpert}, Valencian Region Medical Image Bank [BIMCV] \citep{de2020bimcv}, Brixia \citep{BS-Net2021, borghesi2020covid}, National Institutes of Health [NIH] \citep{wang2017chestx}) can be used for research purposes. Likewise, the datasets for the segmentation tasks (SIIM-ACR pneumothorax segmentation challenge \citep{siim2018pneumothorax}) and the detection (RSNA pneumonia detection challenge \citep{rsna2018pneumonia}) can be used for academic research according to the terms of use.}

\end{document}